\documentclass[review]{elsarticle}
\journal{International Journal of Multiphase Flow}

\usepackage{color}
\usepackage{multirow}
\usepackage{dcolumn}% Align table columns on decimal point
\usepackage{multicol}
\usepackage{graphicx}
\usepackage{caption}
\usepackage{amssymb}
\usepackage{amsmath}
\usepackage{epstopdf}
\usepackage{mathtools}
\usepackage{float}
\usepackage{bm}% bold math

\usepackage{subcaption}
\usepackage{graphicx}
\usepackage{soul}

\biboptions{authoryear}

\biboptions{sort&compress}
\newcommand{\ks}{\textcolor{black}} % for commenting

\begin{document}

\begin{frontmatter}
\title{Jet breakup dynamics of viscoelastic carboxymethyl cellulose solutions}
\author{Ketan Vinayak Warghat$^a$, Yogesh Biswal$^a$, Sukesh Sharma$^a$, Pankaj Sharadchandra Kolhe$^a$, Lakshmana Dora Chandrala$^a$ and \\ Kirti Chandra Sahu$^{b}$\footnote{psk@mae.iith.ac.in, lchandrala@mae.iith.ac.in, ksahu@che.iith.ac.in} }
\address{$^a$Department of Mechanical and Aerospace Engineering, Indian Institute of Technology Hyderabad, Kandi - 502 284, Sangareddy, Telangana, India  \\ 
$^b$Department of Chemical Engineering, Indian Institute of Technology Hyderabad, Kandi - 502 284, Sangareddy, Telangana, India}

\begin{abstract}
We experimentally investigate the breakup dynamics of viscoelastic jets composed of carboxymethyl cellulose (CMC) solutions, focusing on the dripping and Rayleigh regimes at low flow rates. By varying the CMC concentration, needle diameter ($D_n$), and flow rate ($Q$), we analyze the effects of elasticity, viscosity, and flow conditions on jet stability and droplet formation. Our results show that increasing CMC concentration enhances viscoelastic effects, leading to prolonged jet lifetimes, extended liquid threads, and modified pinch-off behavior. At higher concentrations, elasticity suppresses capillary-driven instabilities, slowing thinning and facilitating the formation of beaded structures. We observe that the interplay between inertial, capillary, and elastic forces, influenced by CMC concentration, governs the jet length, droplet volume, and breakup time, with needle diameter and flow rate playing a crucial role in jet breakup phenomenon.
\end{abstract}
\end{frontmatter}


\section{Introduction} \label{sec:intro}

The breakup of liquid jets and droplets is a ubiquitous phenomenon with diverse applications, including combustion \citep{broumand2016liquid}, spray coating \citep{hashemi2023effects}, atomization \citep{ade2022droplet, ade2023droplet}, pharmaceutical manufacturing \citep{mousavi2023comparison, wu2019effects}, and printing technologies \citep{lopez1999break, lopez2004note}. 

Droplet formation, a crucial aspect of jet breakup, involves multiple stages such as jet stretching, neck formation, and pinch-off \citep{borthakur2019dynamics, borthakur2017formation, kirar2024dynamics, sun2018dynamics}. Numerous studies have investigated jet and droplet breakup in Newtonian liquids \citep{ade2024prf, ade2024application}. \citet{birouk2009liquid} reviewed liquid jet disintegration from convergent nozzles, emphasizing the influence of internal flow dynamics and nozzle geometry. \citet{borthakur2019dynamics} conducted numerical simulations of jet formation and breakup from an orifice. They observed that the breakup length increases with the increase in the Weber and Ohnesorge numbers. \citet{wu2019effects} examined the effect of bubbles on the jet breakup, finding that bubbles shorten the breakup length, enhancing atomization efficiency, with lighter gases further amplifying this effect due to momentum conservation. \citet{grant1966newtonian} highlighted the roles of ambient conditions, turbulence, and velocity profiles influencing jet breakup dynamics. \citet{srinivasan2024primary} demonstrated the significant influence of viscosity and fluid rheology on breakup length and jetting transitions. \citet{tanase2023experimental} studied the early stages of liquid injection, focusing on surface tension-driven dynamics during drop formation and pinch-off. Several researchers also investigated the effect of orifice geometry and injection pressure on the jet behavior. \citet{kiaoulias2019evaluation} found that sharp-edge inlets resulted in higher pressure drops and longer breakup lengths in smaller orifices, while chamfered inlets increased breakup lengths in larger orifices. \citet{sharma2014breakup} and \citet{wang2015liquid} found that non-circular orifices (e.g., rectangular, square, triangular) promote greater instability and faster breakup, enhancing atomization. \citet{borthakur2017formation} studied the dynamics of liquid drops and jets during orifice injection, observing self-similar patterns in drop growth and the transition from periodic dripping to elongated jetting. \citet{rajesh2016interfacial} and \citet{rajesh2023drop} found that elliptical and triangular orifices improve the spray characteristics of atomizing jets, resulting in smaller drops compared to circular orifices. \citet{geng2020effect} showed that higher injection pressures and smaller orifice diameters improve atomization. Studies on rectangular and elliptical jets highlighted axis-switching phenomena \citep{morad2020axis, jaberi2019wavelength}, while \citet{kooij2018determines} demonstrated that nozzle geometry and Weber number influence droplet size distribution, with conical nozzles yielding uniform ligaments and flat fan nozzles producing a broader size range.

Based on the aforementioned studies based on Newtonian liquids, jet breakup phenomenon, governed by the interplay of inertial and aerodynamic forces, can be classified into four primary regimes: Rayleigh/dripping, first wind-induced, second wind-induced, and atomization \citep{li2019experimental, chakraborty2022effect, reitz1978atomization, richards1994dynamic, machicoane2023regimes}. The Rayleigh regime involves a continuous jet breaking into droplets due to Plateau-Rayleigh instability, while the dripping regime occurs at very low flow rates, where droplets form directly at the nozzle. As velocity increases, the first wind-induced (convective) regime appears, where aerodynamic forces influence breakup, but the jet remains mostly intact before fragmenting. At higher velocities, the second wind-induced (shear) regime leads to chaotic, non-periodic breakup due to strong aerodynamic shear forces. In the atomization regime, extreme velocities cause complete jet disintegration into a fine spray. The present study investigates the jet breakup dynamics in the Rayleigh and dripping regimes.

A viscoelastic jet, composed of a fluid with both viscous and elastic properties due to polymeric or complex molecular structures, behaves differently from Newtonian jets \citep{li2003drop}. The breakup process of viscoelastic jets is influenced by fluid elasticity, viscosity, surface tension, and flow conditions, making it more complex. The viscoelastic jets exhibit unique characteristics, such as delayed breakup, bead-on-string structures, and intricate filament dynamics, resulting from their ability to store and release elastic energy. \cite{bhat2010formation} showed that viscoelasticity alone cannot produce tiny satellite beads between larger main beads; inertia is essential for their formation. However, viscoelasticity enhances bead growth and delays pinch-off, resulting in longer-lived beaded structures. \citet{cooper2002drop} examined the effect of elasticity on drop formation under gravity using low-viscosity elastic fluids and a Newtonian counterpart. They found that while early-stage necking dynamics were dominated by inertial and capillary forces, independent of elasticity, the break-off process was notably influenced, with higher elasticity resulting in prolonged break-off times and filament lengths due to delayed pinch-off caused by polymer extension in the pinch regions. \citet{mousavi2023comparison} examined the effect of non-Newtonian rheology, particularly shear-thinning behavior, on jet breakup and droplet formation across various regimes. 
Additionally, the secondary breakup of droplets in solutions of deionized (DI) water with varying concentrations of Carboxymethyl Cellulose (CMC) was experimentally investigated by \citet{qian2021experimental}. Using a syringe pump and a needle with a fixed diameter, they observed that the breakup of droplets was influenced by both the Weber and Ohnesorge numbers, which essentially represent the effects of surface tension and viscosity of the liquid, respectively.

As the abovementioned review indicates, only a few studies have systematically examined jet breakup phenomena in viscoelastic jets. While prior research has focused on factors such as needle size and flow rate, further investigations are needed to explore how variations in CMC concentrations and shear-thinning properties affect droplet formation. A more detailed understanding of these dynamics is essential for advancing applications such as inkjet printing, pharmaceuticals, and polymer processing, where precise control over droplet size and breakup behavior is crucial. In the present study, we have systematically conducted experiments to investigate jet breakup dynamics in viscoelastic liquids, specifically solutions of deionized water with varying concentrations of Carboxymethyl Cellulose (CMC), using shadowgraphy and high-speed imaging techniques. We also varied the size of the needle to explore the influence of flow rate on the breakup dynamics. Our analysis specifically focuses on intriguing phenomena in viscoelastic jets, such as the jet breakup length and the upward motion of satellite droplets, which are driven by the elastic stress in CMC solutions.

The rest of the manuscript is organized as follows: Section \ref{sec:exp} provides a detailed description of the experimental methodology, including setup details, varied parameters, data acquisition techniques, post-processing methods, and jet characterization. Section \ref{sec:dis} presents the discussion of the results. Finally, Section \ref{sec:conc} offers the concluding remarks.

\section{Experimental setup and procedure} \label{sec:exp}

\begin{figure}
\centering
\includegraphics[width=1.0\textwidth]{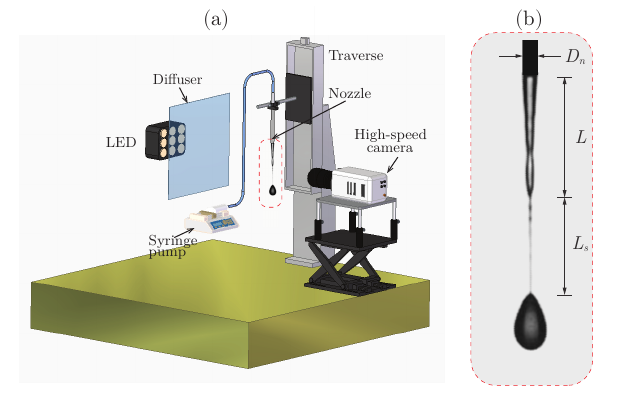}
\caption{(a) A schematic representation of the experimental setup, illustrating the dispensing mechanism for the CMC solution, which is controlled by a syringe pump, along with the shadowgraph-based image-capturing system. (b) A schematic diagram showing the needle and jet breakup length, where $D_n$ is the needle diameter, $L$ is the jet length, and $L_s$ is the liquid thread length.}
\label{fig:exp_setup}
\end{figure}

We experimentally investigate the jet breakup dynamics of viscoelastic carboxymethyl cellulose (CMC) solutions in the Rayleigh/dripping regime. Figure \ref{fig:exp_setup}(a) presents a schematic of the experimental setup, which consists of a syringe pump connected to a dispensing needle and a shadowgraph system. The shadowgraph system includes a high-speed camera and a diffuser sheet to uniformly disperse light from an LED source. We consider three deionized water (DI) and CMC solutions by varying the CMC concentration to 1.0 wt.\%, 2.0 wt.\%, and 3.0 wt.\%. Additionally, the inner diameter of the blunt needles ($D_n$) and the flow rate ($Q$) are varied in our experiments. Specifically, we use needle diameters of $D_n = 0.5$ mm, 1.0 mm, and 2.0 mm, while the flow rate is varied from $Q = 2 \times 10^{-8}$ m$^3$/s to $Q = 14 \times 10^{-8}$ m$^3$/s. 

\ks{A food-grade carboxymethyl cellulose (CMC) powder (molecular weight $\approx$ 90 kDa; degree of substitution, DS = 0.7), obtained from a local manufacturer, was used in the experiments. The CMC solutions were prepared by mixing the CMC powder with deionized (DI) water in specified weight ratios. The mixture was manually stirred for 30 minutes to ensure homogeneity. Subsequently, it was placed in a vacuum desiccator to eliminate any entrapped air. A positive displacement syringe pump, calibrated using standard syringe volumes, was employed to maintain a constant and uniform flow rate throughout the experiments. During the entire procedure, the temperature was maintained at room temperature ($25^\circ$C).} Fresh solutions are prepared for each experiment to maintain consistent solution properties, as they can change over time. A new syringe is used for every trial to ensure proper flow and prevent cross-contamination. The syringe pump, assisted by a computerized controller, maintains a constant flow rate at the exit of the needle, ensuring a controlled and steady liquid stream.

 The solution was prepared by manual stirring at room temperature to minimize air bubble formation. Subsequently, it was placed in a vacuum desiccator to eliminate any entrapped air. A positive displacement syringe pump, calibrated with standard syringe volumes, was employed to ensure a constant and uniform flow rate throughout the experiments.

\begin{figure}
\centering
\hspace{0.5cm} {\large (a)} \hspace{5.5cm} {\large (b)}\\
\includegraphics[width=0.45\textwidth]{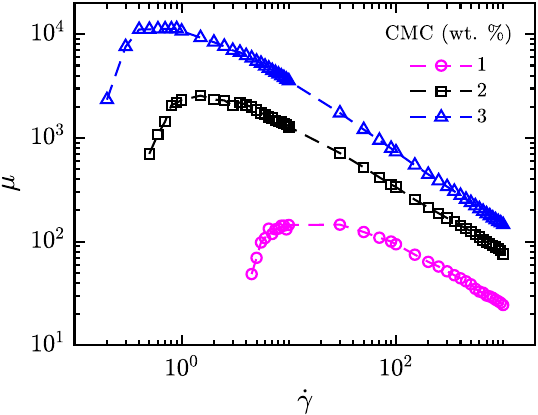} \hspace{0.5cm}
\includegraphics[width=0.45\textwidth]{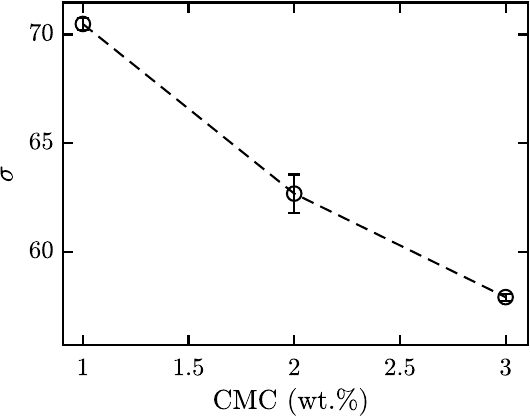}
\caption{(a) Variation of the dynamic viscosity $\mu$ (mPa$\cdot$s) with the shear rate $\dot{\gamma}$ for different deionized water and CMC solutions with varying CMC concentration (wt. \%). (b) Variation of the surface tension $\sigma$ (mN/m) of the solution with the CMC concentration (wt. \%).}
\label{fig:fig2}
\end{figure}

First, we characterize the fluid properties of the CMC solutions. The density of the CMC solutions is measured using a hydrometer, where each solution is filled to a specific level to determine its volume, and the mass is obtained using a weighing balance. This allows us to calculate the density of solutions with varying CMC concentrations. We found that the density of the solutions remains approximately constant at $1066 \pm 5$ kg/m$^3$. The viscosity of the solution is measured using a Rheometer (RheolabQC SN82215842, Make: Anton Paar). The RheolabQC is a rotational rheometer operating based on the Searle principle, with speed regulation controlled by an electric motor. It measures viscosity as a function of shear rate. Figure \ref{fig:fig2}(a) shows the viscosity versus shear rate relationship on a log-log scale for different concentrations of the CMC solution. The simplified Carreau-Yasuda rheological model is used to derive the rheological properties of the solutions, which is given by \citep{Cross,premlata2017numerical}
\begin{equation} 
\mu = \mu_0 \left[ 1 + (\lambda \Pi)^2 \right]^{\frac{(n-1)}{2}}.
\label{Carreau-Yasuda} 
\end{equation}
Here, $\mu_0$ represents the dynamic viscosity at zero shear rate, $\Pi$ is the second invariant of the strain rate tensor, $n$ is the flow index, and $\lambda$ is the characteristic time constant. The rheological properties of the solutions are listed in Table \ref{Properties}. The results indicate that increasing CMC concentration increases the zero-shear viscosity ($\mu_0$) and the characteristic relaxation time ($\lambda$). The viscosity of the deionized water is 1 mPa$\cdot$s. For a 1.0\% CMC solution, the value of the zero-shear viscosity ($\mu_0$) is 158.0 mPa$\cdot$s and the relaxation time is 0.021 s. For 2.0\% and 3.0\% CMC solutions, the viscosity rises to 2129.7 mPa$\cdot$s and 3551.8 mPa$\cdot$s, respectively, while the relaxation time increases to 0.191 s and 0.314 s. Additionally, the power-law index ($n$), which reflects the extent of shear-thinning behavior, decreases with the increase in the CMC concentration. For the 1.0\%, 2.0\%, and 3.0\% CMC solutions, the values of $n$ are 0.398, 0.374, and 0.286, respectively. This indicates \ks{an increase} in shear-thinning behavior as the CMC concentration increases. The surface tension of the solution is measured using a Drop Shape Analyzer (DSA25, KRÜSS) via the pendant drop method. The system features software-controlled dosing, high-power monochromatic LED illumination, and a zoom-lens camera. \ks{In our experiments, the time between droplet formation, filament formation, breakup, and subsequent droplet generation is relatively short. Therefore, dynamic surface tension was measured by analyzing the droplet shape at a fixed interval of 60 seconds after its formation.} As shown in Figure \ref{fig:fig2}(b), the surface tension ($\sigma$) of the CMC solution decreases with increasing CMC concentration. 

%2

\begin{table}
\centering
\begin{tabular}{cccc}
\hline
CMC ($\%$) & $\mu_0$ (mPa. s) & $\lambda$ (s) & $n$\\ \hline
0 (deionized water)     & 1.0 & 0.0 & 1.0 \\
1.0     & 158.0 & 0.021 & 0.398\\
2.0     & 2129.7 & 0.191 & 0.374\\
3.0     & 3551.8 & 0.314 & 0.286 \\ \hline
\end{tabular}
\caption{The rheological properties of the CMC solutions derived from the Carreau-Yasuda model (Eq. \ref{Carreau-Yasuda}).}
\label{Properties}
\end{table}

\begin{figure}
\centering
\includegraphics[width=0.4\textwidth]{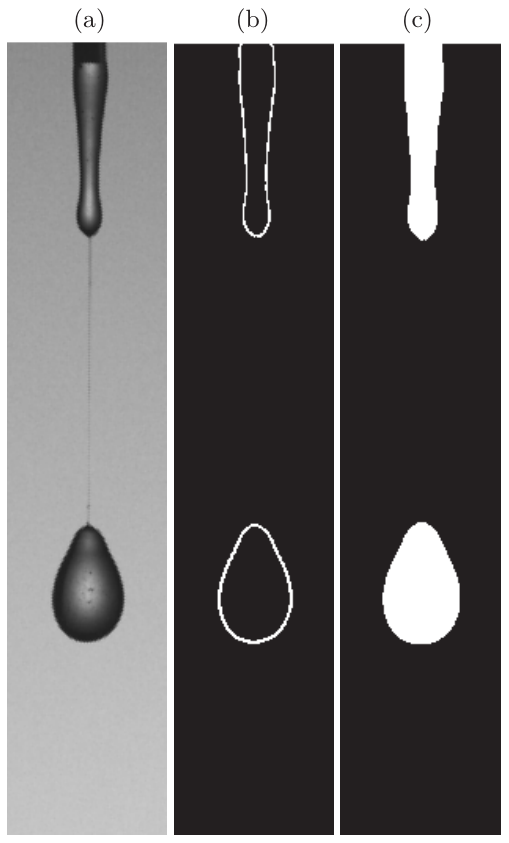}
\caption{Various image processing steps for droplet boundary detection.}
\label{fig:image}
\end{figure}

The orientation of the dispensing nozzle, traversing system, and camera arrangement are shown in figure \ref{fig:exp_setup}(a). A high-speed shadowgraphy technique using a diffused LED light source is employed to visualize liquid jets. The high-speed camera (Model: Phantom VEO-710L) equipped with a 100 mm macro lens, records droplet formation at 15,000 frames per second (fps) with a resolution of $512 \times 512$ pixels. The aperture is adjusted based on the light source and needle position to optimize image quality. The droplets are analyzed at various points along the length of the jet. Figure \ref{fig:exp_setup}(b) provides a schematic diagram showing the needle and jet breakup length, where $D_n$ denotes the needle diameter, $L$ represents the jet length, and $L_s$ corresponds to the liquid thread length. The image processing of the recorded video of the jets is performed using the ImageJ and \textsc{Matlab}$^{\circledR}$ software. The videos captured with Phantom PCC software are converted into individual frames for analysis. The images are cropped and binarized in ImageJ before further processing in \textsc{Matlab}$^{\circledR}$ to estimate various physical dimensions of the jets with different CMC concentrations. \ks{To compute the droplet volume from a series of images, we implement a structured image processing pipeline. The first step involves extracting object boundaries from raw images using the Canny edge detection algorithm, which includes Gaussian filtering to reduce noise, gradient computation, edge thresholding, and edge tracking (see, \citet{canny1986computational} for details). After generating the edge map, we apply a morphological filling operation to identify and fill closed regions enclosed by the detected boundaries. This step transforms the enclosed areas, typically appearing as hollow contours, into solid regions by filling them with white pixels. The various image processing steps are illustrated in Figure \ref{fig:image}.}

%3

\begin{figure}
\centering
\includegraphics[width=0.4\textwidth]{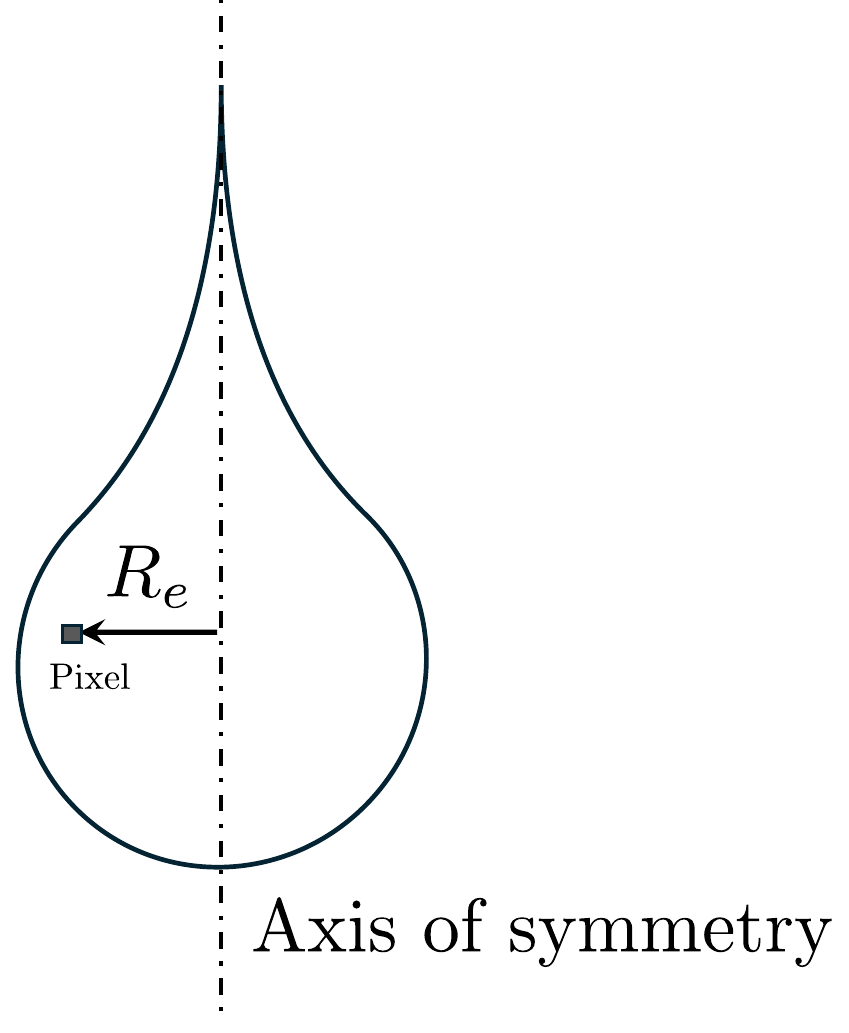}
\caption{A schematic representation of the droplet volume calculation. The droplet profile is discretized into individual pixel elements. Each pixel area $A_{e,i}$ is revolved around the axis of symmetry using its corresponding equivalent radius $R_{e,i}$ to estimate the total volume.}
\label{fig:dropvol}
\end{figure}

\ks{Further, we observe that the elongated droplets do not conform to any standard geometric shape. Therefore, we adopt an alternative method to estimate the droplet volume, as detailed below. The volume is calculated by determining the pixel area and revolving it about the axis of symmetry, using an equivalent radius $R_{e,i}$ corresponding to each pixel area $A_{e,i}$ enclosed within the droplet (see, figure \ref{fig:dropvol}). The total droplet volume is then computed using the following expression:
\begin{equation}
V_d={1 \over 2} \sum_{i=1}^N (2 \pi R_{e,i} A_{e,i}),
\end{equation}
where $N$ is the total number of pixels.}

%4

To gain a deeper understanding of jet behavior, the space-time evolution of the jet is analyzed at different axial positions using a stack of 30,000 frames captured over a 2-second duration. Additionally, droplet morphology is examined at various distances from the needle tip to evaluate the influence of thread length variations and solution concentrations. \ks{The results reported in the following section are obtained from three repetitions for each set of parameters.}

\section{Results and discussion} \label{sec:dis}

\begin{figure}
\centering
\includegraphics[width=0.95\textwidth]{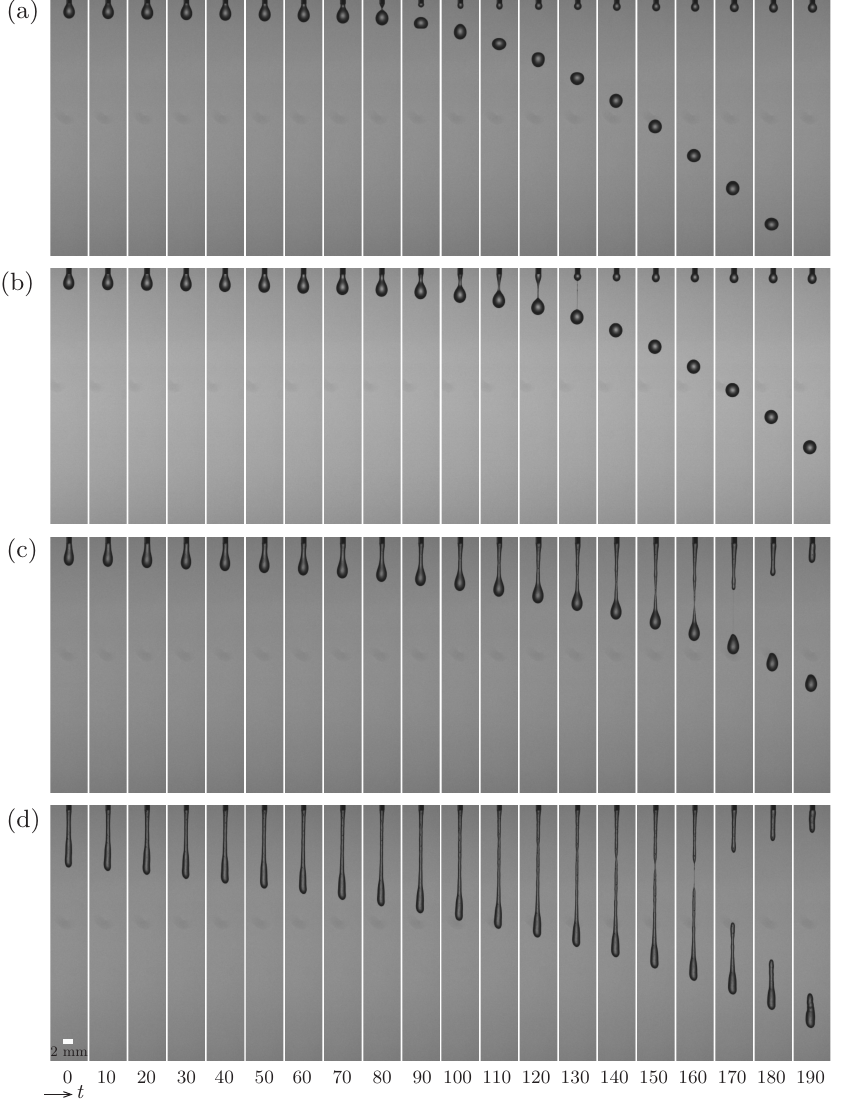}
\caption{Temporal evolution of jet breakup dynamics for different deionized water and CMC solutions. (a) Deionized water without CMC, and deionized water and CMC solution with (b) $1.0$ wt.\%, (c) $2.0$ wt.\%, and (d) $3.0$ wt.\% CMC. Here, the flow rate ($Q$) and needle diameter ($D_n$) are fixed at $Q = 8.334 \times 10^{-8}$ m$^3$/s and $D_n = 1$ mm. The time $(t)$ in seconds is shown at the bottom of each panel, with $t=0$ representing the moment when the droplet begins to emerge from the nozzle. A scale bar is shown in the bottom-left corner of panel (d).}
\label{fig:shadowgraphy}
\end{figure}

We begin the presentation of our results by analyzing the jet breakup dynamics for different deionized water and CMC solutions. Figure \ref{fig:shadowgraphy}(a-d) shows the temporal evolution of jet breakup dynamics for (a) deionized water and CMC solutions with (b) 1.0 wt.\%, (c) 2.0 wt.\%, and (d) 3.0 wt.\% CMC, respectively. Here, flow rate ($Q$) and needle diameter ($D_n$) are fixed at $Q = 8.334 \times 10^{-8}$ m$^3$/s and $D_n = 1$ mm.  As shown in Figure \ref{fig:shadowgraphy}(a), in the absence of CMC, the jet exhibits a dripping regime, which typically occurs at lower liquid flow rates, where surface tension forces dominate over inertial forces \citep{borthakur2017formation}. In this regime, droplet formation takes place near the nozzle exit, resulting in distinct and relatively uniform droplets. The breakup process is primarily governed by capillary action, which arises from the interfacial tension between the liquid and the surrounding medium. At a later stage, it can be observed that as the droplet is released from the nozzle, it undergoes shape oscillations due to the interplay of surface tension, gravity, and viscous forces as it falls downward \citep{balla2019shape,agrawal2017nonspherical}.

%5

For deionized water and CMC solution with $1.0$ wt.\% CMC, the liquid accumulates at the nozzle tip to form a droplet, which is then pushed downward by the combined effects of gravity and pressure forces (Figure \ref{fig:shadowgraphy}b). The rheological parameters of this solution are $\mu_0 = 158$ mPa·s, $\lambda = 0.021$, and $n = 0.398$. The parameter $\lambda$ represents a characteristic time constant that defines the transition from Newtonian to shear-thinning behavior and plays a crucial role in determining how viscosity changes with shear rate. It specifically influences the balance between capillary-driven thinning and polymeric stress relaxation, which ultimately affects droplet size and the formation of satellite droplets. In jet breakup studies, viscoelastic fluids exhibit complex thinning and droplet detachment dynamics. Figure \ref{fig:shadowgraphy}(b) shows that a thin thread-like structure initially attaches the droplet to the nozzle (see, $100 \le t \le 130$). As time progresses, the thread elongates, thins, and the droplet eventually detaches. A close inspection of Figure \ref{fig:shadowgraphy}(b) reveals that the detachment time of the droplet is approximately 140 seconds, significantly longer than the detachment time observed for pure water ($t \approx 90$ s). Additionally, due to the dominant effects of surface tension and shear-thinning behavior, the droplet exhibits smaller shape oscillations (remaining mostly spherical) after detaching from the nozzle compared to those observed in the case of pure water (Figure \ref{fig:shadowgraphy}a).

\begin{table}
\centering
\begin{tabular}{ccc}
\hline
CMC (\%) & $We (= \rho V^2 D_n/\sigma)$ & $Oh (= \mu/\sqrt{\rho \sigma D_n})$\\ \hline
1.0 & 0.17 & 578\\ 
2.0 & 0.19 & 8239\\ 
3.0 & 0.2 & 14295 \\ \hline
\end{tabular}
\caption{\ks{The values of the Weber number ($We = \rho V^2 D_n/\sigma$) and Ohnesorge number ($Oh = \mu/\sqrt{\rho \sigma D_n}$) for $Q = 8.334 \times 10^{-8}$ m$^3$/s and $D_n = 1$ mm. Here, $\rho$ denotes the density of the solution, $\mu$ is the dynamic viscosity, $\sigma$ is the surface tension, and $V$ is the average velocity of the jet.}}
\label{Table_T2}
\end{table}

An increase in the CMC concentration (wt.\%) in the solution further complicates the dynamics by transitioning the breakup mechanism from the dripping regime to the Rayleigh regime. \ks{Table \ref{Table_T2} presents the values of the Weber number $(We = \rho V^2 D_n/\sigma)$ and the Ohnesorge number $(Oh = \mu/\sqrt{\rho \sigma D_n})$ corresponding to a representative flow rate and nozzle diameter ($Q = 8.334 \times 10^{-8}$ m$^3$/s and $D_n = 1$ mm). It is observed that increasing the concentration of CMC in the solution does not significantly affect the Weber number, whereas the Ohnesorge number increases markedly. This suggests that the influence of viscosity becomes increasingly dominant at higher CMC concentrations. Thus, the breakup phenomenon observed at low CMC concentrations corresponds to the dripping regime $(Oh \sim 500)$, whereas at higher concentrations, it transitions into the Rayleigh breakup regime $(Oh \sim 10^4)$.} As shown in Figures \ref{fig:shadowgraphy}(c) and \ref{fig:shadowgraphy}(d) for 2.0 wt.\% CMC and 3.0 wt.\% CMC, the thread thickness increases, forming a jet (continuous liquid flow), and the detachment time of the droplet also increases as we increase the CMC concentration. Moreover, due to the substantial increase in viscosity (see Table \ref{Carreau-Yasuda}), the liquid blob detached from the jet no longer assumes a spherical shape. It is important to note that after the drop detaches, the elongated thread gives rise to a small spherical satellite drop, in addition to the larger drop or ligament detached from the main jet. This satellite drop exhibits an upward motion because of the elastic nature of the solution, which becomes more pronounced as the CMC concentration increases. This phenomenon will be explored in the following section.

%6
\begin{figure}
\centering
\includegraphics[width=0.95\textwidth]{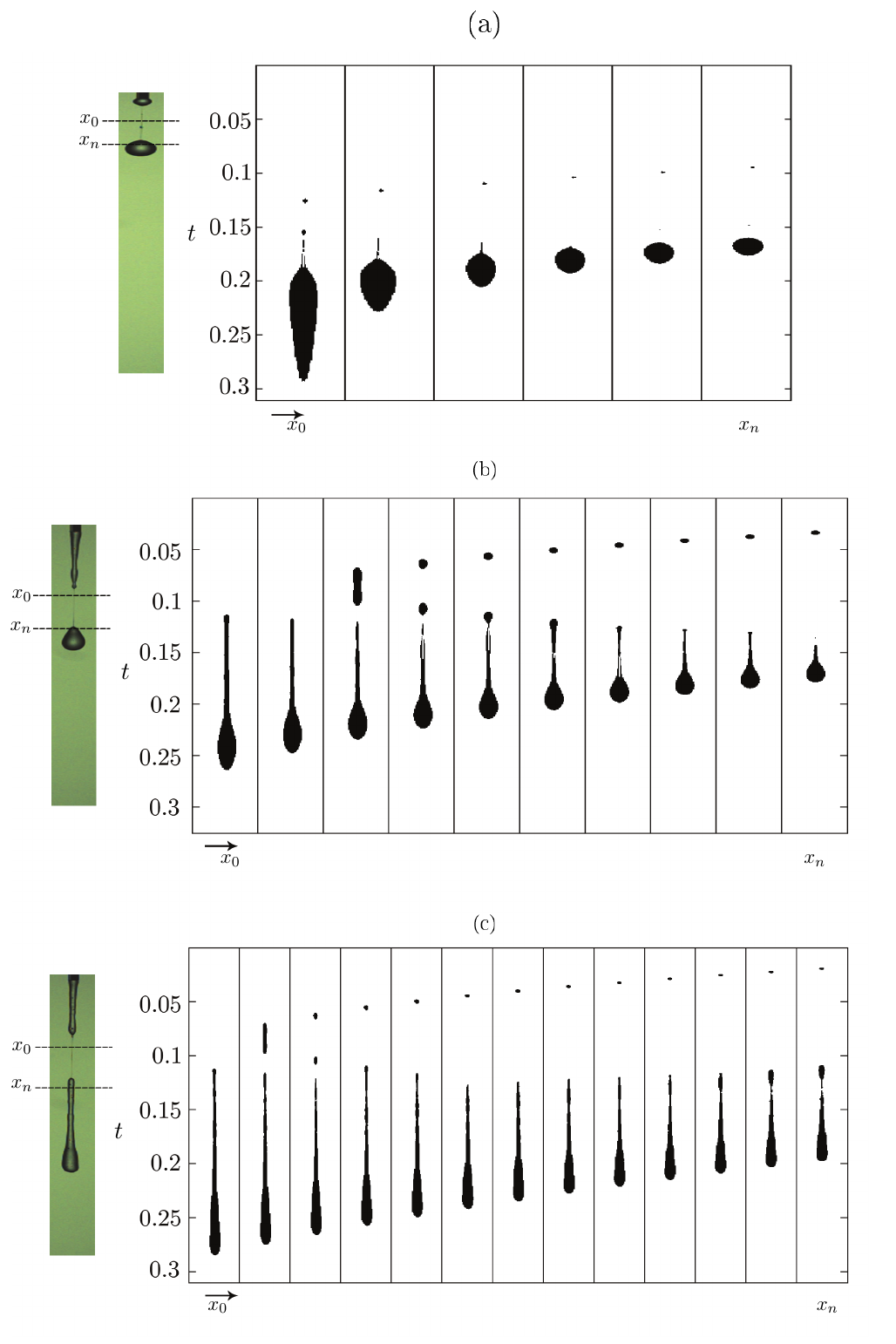}
\caption{Spatio-temporal evolution of a droplet at a symmetric location across different vertical step positions for deionized water and CMC solutions with varying concentrations: (a) $1.0$ wt.\% CMC, (b) $2.0$ wt.\% CMC, and (c) $3.0$ wt.\% CMC.}
\label{fig:space_time}
\end{figure}

Figure \ref{fig:space_time} presents the space-time evolution of the droplet at different axial locations ($x_{0}$ to $x_{n}$ as indicated in the inset) for varying CMC concentrations, highlighting the influence of viscoelasticity on droplet dynamics. It can be seen in Figure \ref{fig:space_time} that, for the 1\% CMC solution, the droplet undergoes elongation, leading to the formation of secondary and satellite droplets following the detachment of the liquid thread. The elastic nature of the fluid generates a restoring force, pulling the secondary droplet upward, which is a signature of elastic recoil commonly observed in viscoelastic fluids \citep{cooper2002drop,smolka2003drop}.  The satellite drop integrates and merges with the primary drop due to the capillary action. As the CMC concentration increases to 2\%, the breakup dynamics become more complex, with the secondary droplet further fragmenting into two smaller droplets (Figure \ref{fig:space_time}b). One of these droplets merges back with the primary droplet due to capillary forces, while the other moves upward, driven by the elastic tension in the thinning filament. Furthermore, the string of the primary droplet eventually coalesces into a single drop, demonstrating the competing effects of elasticity and surface tension. At 3\% CMC concentration, the overall trend remains similar to the 2\% case, but with a significantly longer string length before detachment (Figure \ref{fig:space_time}c). This suggests that higher CMC concentrations enhance viscoelastic stretching, delaying the breakup process and promoting the formation of elongated liquid filaments. The interplay between elasticity, capillary forces, and inertial effects becomes increasingly evident, as higher polymer concentrations resist rapid thinning, leading to prolonged filament lifetimes and modified droplet coalescence dynamics. The time scale on the vertical axis in Figure \ref{fig:space_time} supports viscosity measurements and variations in $\lambda$. The droplet volume, and consequently its inertia, influences the shear rate in the experiments. Thus, increased inertia in 3\% CMC leads to a higher shear rate, consistent with the droplet volume measurements discussed later.

%7
\begin{figure}
\centering
\includegraphics[width=0.6\textwidth]{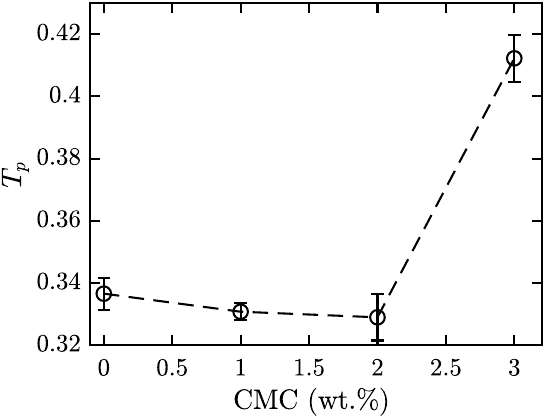}
\caption{Variation of the time period ($T_p$) of droplet occurrence with the CMC (wt. \%) concentration. Here, for all values of the CMC (wt. \%) concentration, the flow rate $(Q)$ and needle diameter $(D_n)$ are fixed at $Q = 8.334 \times 10^{-8}$ m$^3$/s and $D_n = 1$ mm. \ks{The error bars represent the standard deviation obtained from three repetitions under identical conditions.}}
\label{fig:800spray}
\end{figure}

Figure \ref{fig:800spray} depicts the time period $(T_p)$ of primary droplet occurrence, obtained from space-time images (figure \ref{fig:space_time}), for a flow rate of $Q = 8.334 \times 10^{-8}$ m$^3$/s and a needle diameter of $D_n = 1$ mm. The results show that the time period decreases as the CMC concentration reaches 2\%. However, with a further increase in CMC concentration, the time period increases due to the higher viscosity of the solution, which thickens the fluid and extends the droplet formation time for the given flow rate and needle size.

%8
\begin{figure}
\centering
\hspace{0.5cm} {\large (a)} \hspace{5.5cm} {\large (b)}\\
\includegraphics[width=0.45\textwidth]{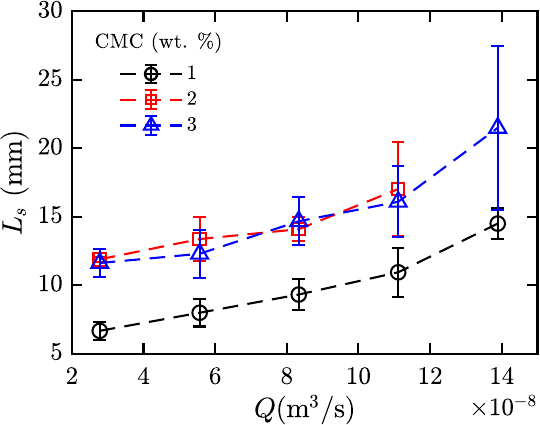} \hspace{2mm} 
\includegraphics[width=0.45\textwidth]{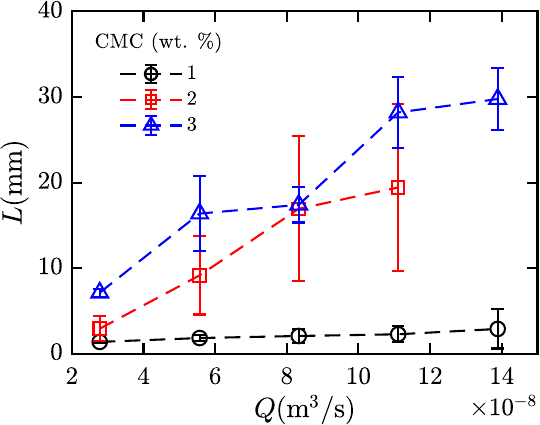} \\
\hspace{0.6cm} {\large (c)} \\
\includegraphics[width=0.45\textwidth]{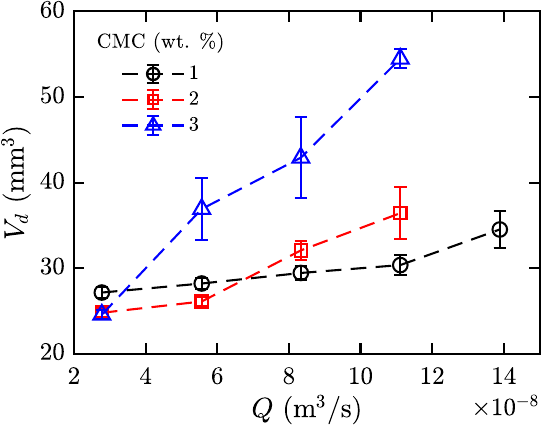}
\caption{Variation of the (a) liquid thread length $(L_s)$, (b) jet length $(L)$, and (c) drop volume $(V_d)$ with the flow rate ($Q$ in m$^3$/s) for different CMC concentrations (wt. \%). Here, $D_n = 1$ mm. The liquid thread length $(L_s)$ and jet length $(L)$ are schematically shown in Figure \ref{fig:exp_setup}(b). \ks{The error bars represent the standard deviation obtained from three repetitions under identical conditions.}}
\label{fig:fig6}
\end{figure}

Next, we conduct a quantitative analysis of the physical parameters associated with the jet breakup process, as illustrated in Figure \ref{fig:exp_setup}. Figure \ref{fig:fig6}(a-c) depicts the variation of the liquid thread length $(L_s)$, jet length $(L)$, and drop volume $(V_d)$ with the flow rate ($Q$ in m$^3$/s) for different CMC concentrations (wt. \%). Here, the diameter of the needle is fixed at $D_n = 1$ mm. It can be seen that as the flow rate increases, the liquid thread length ($L_s$) increases due to the higher momentum in the penetrating fluid. However, the presence of viscoelasticity, introduced by the addition of CMC, modifies this behavior by delaying capillary breakup. At higher polymer concentrations, elastic stresses resist deformation, suppressing Rayleigh-Plateau instabilities and leading to a longer liquid thread. As the jet undergoes the dripping regime, the jet length ($L$) is zero for the Newtonian fluid at the flow rates considered in the present study. For low CMC concentration (1 wt. \% CMC), the jet length ($L$) increases slightly (but remains very small) with the increase in the flow rates. For intermediate CMC concentration (2 wt. \% CMC), the jet length ($L$) increases with flow rate, primarily due to inertial stretching. However, for higher CMC concentration (3 wt. \% CMC), the elastic stresses counteract elongation, leading to a more stable and longer jet compared to Newtonian fluids. The droplet volume ($V_d$) also increases with flow rate, as more fluid is ejected per unit time. However, at higher CMC concentrations, viscoelastic effects alter the pinch-off dynamics, delaying droplet detachment and resulting in larger or more elongated droplets. These observations highlight the complex interplay between inertia, capillarity, and viscoelastic resistance in determining jet stability and drop formation.

%9
\begin{figure}
\centering
\hspace{0.5cm} {\large (a)} \hspace{5.5cm} {\large (b)}\\
\includegraphics[width=0.45\textwidth]{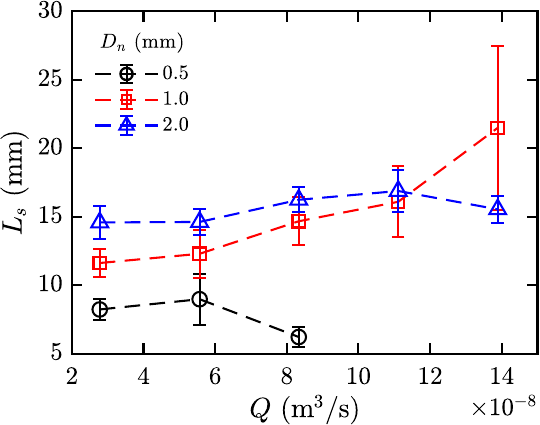} \hspace{2mm} 
\includegraphics[width=0.45\textwidth]{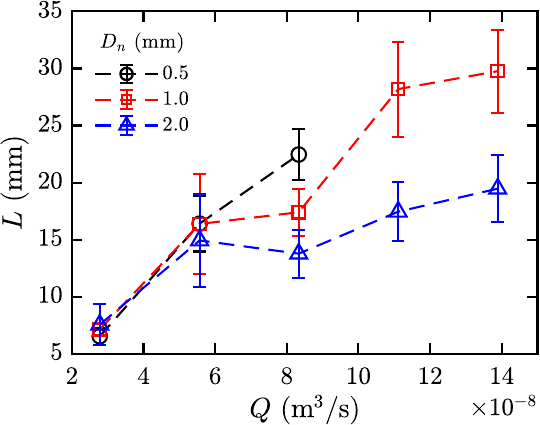} \\
\hspace{0.6cm} {\large (c)} \\
\includegraphics[width=0.45\textwidth]{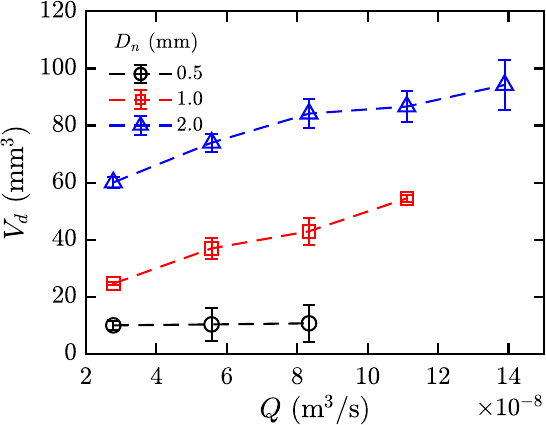}
\caption{Variation of the (a) liquid thread length $(L_s)$, (b) the jet length $(L)$, and (c) drop volume $(V_d)$ with the flow rate ($Q$ in m$^3$/s) for different needle diameter $(D_n)$. Here, the CMC concentration is maintained at 3 wt.\%. \ks{The error bars represent the standard deviation obtained from three repetitions under identical conditions.}}
\label{fig:fig7}
\end{figure}

Figure \ref{fig:fig7} illustrates the variation of liquid thread length ($L_s$), jet length ($L$), and droplet volume ($V_d$) with flow rate ($Q$) for different needle diameters ($D_n$) while maintaining a fixed CMC concentration of 3 wt.\%. As the flow rate increases, the liquid thread length ($L_s$) extends due to the increased fluid momentum. However, the needle diameter ($D_n$) significantly influences thread stability. A larger needle diameter produces a thicker jet, which is more resistant to capillary instabilities and thus extends further before breaking up. In contrast, a smaller needle diameter results in a thinner jet where surface tension forces dominate over inertia, leading to an earlier breakup. For viscoelastic fluids like CMC solutions, polymeric stresses resist thinning and delay breakup, further extending the liquid thread compared to Newtonian fluids. Similarly, the jet length ($L$) increases with flow rate due to inertial stretching, but its stability is affected by $D_n$. A larger needle diameter generates a thicker, more stable jet, reducing the impact of capillary-driven instabilities and leading to a longer jet. Conversely, a smaller needle diameter forms a thinner jet that is more susceptible to breakup, resulting in a shorter jet length. Additionally, viscoelastic effects enhance jet stability by introducing elastic stresses that resist capillary breakup, allowing the jet to extend further before destabilizing. The droplet volume ($V_d$) also increases with flow rate since a higher $Q$ results in a larger volume of fluid being ejected per unit time. The needle diameter plays a crucial role in droplet formation, with larger needle diameters producing bigger droplets due to the increased volume of ejected fluid before pinch-off. Conversely, smaller needle diameters generate smaller droplets, as surface tension forces dominate the pinch-off process. Furthermore, viscoelasticity alters pinch-off dynamics by delaying droplet detachment, leading to larger and more elongated droplets compared to Newtonian fluids. These observations highlight the interplay of inertia, capillarity, and viscoelastic stresses in governing jet stability and droplet formation, with needle diameter serving as a critical parameter in controlling these processes.

\section{Concluding remarks} \label{sec:conc}

We systematically conduct experiments to investigate the breakup dynamics of viscoelastic jets composed of carboxymethyl cellulose (CMC) solutions, focusing on the dripping and Rayleigh regimes at low flow rates. The study explores the effects of elasticity, viscosity, and flow conditions on jet stability and droplet formation by varying the CMC concentration, the inner diameter of the blunt needles ($D_n$), and the flow rate ($Q$). Utilizing high-speed shadowgraphy and image analysis, we quantify the jet characteristics, including delayed breakup, elongated liquid threads, and the formation of secondary droplets exhibiting upward motion due to elastic stresses. Our results reveal that increasing CMC concentration amplifies viscoelastic effects, resulting in prolonged jet lifetimes, extended liquid threads, and altered pinch-off dynamics. At higher concentrations, elasticity counteracts capillary-driven instabilities, suppressing rapid thinning and facilitating the formation of beaded structures. Furthermore, our quantitative analysis demonstrates that the interplay between inertial, capillary, and elastic forces governs the jet length, droplet volume, and breakup time. While lower CMC concentrations exhibit inertial stretching, higher concentrations introduce significant viscoelastic resistance, stabilizing the jet and changing droplet coalescence dynamics. The impact of needle diameter and flow rate variations further highlights the sensitivity of jet breakup to external parameters. 

Overall, this study provides valuable insights into the complex physics of viscoelastic jet breakup, contributing to a deeper understanding of fluid fragmentation in non-Newtonian systems. These findings have implications for various industrial applications, including inkjet printing, pharmaceutical droplet formation, and polymer processing, where precise control over droplet size and stability is crucial. Future work could explore the role of additional rheological properties, such as extensional viscosity, and investigate jet breakup in turbulent or multiphase environments to further enhance our understanding of viscoelastic fluid dynamics.

%\vspace{2mm}
%\noindent{\bf Declaration of Interests:} The authors report no conflict of interest. \\
%\\
%\noindent{\bf Acknowledgement:} {K.C.S. thanks the IIT Hyderabad for their financial support provided through grant IITH/CHE/F011/SOCH1.}

%\bibliography{bibl}

\end{document}